\documentstyle[preprint, pre, aps]{revtex}
\begin{document} 
\draft 
\title{Comment on ``Ehrenfest times for classically chaotic systems''}
\author{Steven Tomsovic$^1$ and Eric J.~Heller$^2$}
\address{$^1$Department of Physics, Washington State University, Pullman,
WA 99164-2814}
\address{$^2$Department of Physics and Department of Chemistry and
Chemical Biology, Harvard University, Cambridge, MA 02138}
\date{\today}
\maketitle

In a recent Rapid Communication~\cite{silvestrov}, the authors, Silvestrov
and Beenakker, introduce a way to lengthen the Ehrenfest time, $\tau$,
for fully chaotic systems.  We disagree with several statements made
in their paper, and address the following points essential to their
conclusions: 1) it is not true that all semiclassical approximations for
chaotic systems fail at a so-called `logtime', $\tau \propto - \ln
(\hbar)$, differing only by a numerical coefficient; and 2) the
limitation of the semiclassical approximation as expressed in the
authors' Eq.~(8) is not limited by their argument leading to Eq.~(12).

It is important to distinguish between the correspondence of quantum and
classical dynamical propagations, and the faithfulness of semiclassical
approximations.  If one takes the Ehrenfest time, $\tau$, to be the upper
limit for which a quantum mechanical wave packet is described by solving
classical equations of motion without invoking a semiclassical
construction of the wave packet, then the Ehrenfest time increases
logarithmically slowly for chaotic systems as $\tau \propto \lambda^{-1}
\ln (S/\hbar)$~\cite{zaslavsky,bbtv}; there is no controversy on this
point.  In this expression $\lambda$ is sum of the positive Lyapunov
exponents, and $S$ is some characteristic classical action such as that
of the shortest periodic orbit.  If, instead, one defines $\tau$ as the
time scale beyond which the semiclassical approximation no longer
faithfully reproduces the quantum propagation of a wave packet, then
$\tau$ is not a so-called ``logtime'', but is proportional to inverse
algebraic powers of $\hbar$~\cite{stadium,kickedrotor,bakersmap}.  

The precise exponent in the breakdown time scale has been shown to depend
on a few basic features of the chaotic dynamical system being
considered.  We mention work on three separate paradigms of chaos.  It
was shown in the stadium billiard~\cite{stadium}, that $\tau \propto
\hbar^{-1/2} \ln {S/\hbar}$ (essentially $\hbar^{-1/2}$).  The
$\hbar^{-1/2}$ behavior was linked to the fact that the stable and
unstable manifolds associated with trajectories in the stadium have
discontinuities in their slopes where they fold over upon themselves. 
The $\ln {S/\hbar}$ part of the expression is due to the `stickiness' of
phase space in the neighborhood of the marginally stable bouncing ball
trajectories.  In contrast, a general dynamical system possessing stable
and unstable manifolds that are continuous in their slopes gives
$\tau \propto \hbar^{-1/3}$~\cite{kickedrotor}; this was illustrated with
the kicked rotor.  A third example that has been studied extensively is
the quantum bakers map.  There it was shown that for some quantities, the
breakdown time scale could be as great as $\tau \propto
\hbar^{-1}$~\cite{bakersmap}, although $\hbar^{-1/2}$ was
typical~\cite{smilansky}.  

Note that the semiclassical approximations in
Refs.~\cite{stadium,kickedrotor,bakersmap,smilansky} involve no
uniformizations or caustic corrections.  They are, in fact, either
exactly or poor man's versions of the standard WKB method, and developed
specifically for chaotic systems.  For wave packets, the standard
time-dependent WKB method involves sets of complex
trajectories~\cite{wpd}.  Nevertheless, in the above cited work on
$\tau$, no classically non-allowed processes are taken into account.  One
essential ingredient relied upon in these works, the `area-$\hbar$ rule'
is contained in Ref.~\cite{bbtv}.  This rule is in contradiction with the
argument of Ref.~\cite{silvestrov} leading to Eq.~(12) which contains the
relation $\hbar^{7/6-c}<<1$, $c$ being the coefficient of proportionality
in the logtime scale relation.  The consequences of the area-$\hbar$ rule
carefully considered in conjunction with the geometrical properties of
evolving stable and unstable manifolds give a precise formulation of the
semiclassical breakdown due to caustics and the resultant algebraic time
scales~\cite{stadium,kickedrotor,bakersmap}.  The crucial point is that
the {\it distance} between local classical manifolds (the criterion used
by Silvestrov and Beenakker) is actually of no importance - what matters
is the {\it area} enclosed by following the manifold from one branch to
the next in a given locality.  To miss this point unfortunately leads to a
qualitatively different and incorrect result.  

Finally, we do agree with the authors that there should be
important or morphological distinctions in the nature of evolving wave
packets as they surpass each relevant time scale.  Examples include
interference phenomena necessarily arising beyond the logtime, and
localization effects sometimes supressing classical diffusion beyond
algebraic time scales~\cite{grempel}.

\end{document}